\documentclass[aps,showpacs,preprintnumbers,amsmath, amssymb]{revtex4}

\usepackage{float}
\usepackage{graphics,epsfig}
\usepackage{graphicx}
\usepackage{dcolumn}
\usepackage{bm}

\begin{document}
\baselineskip=0.8 cm
\title{\bf Improved parametrization of the growth index for dark energy and DGP models}
\author{Jiliang Jing }
\email{jljing@hunnu.edu.cn}
 \affiliation{Institute of Physics and
Department of Physics,
Hunan Normal University,  Changsha, Hunan 410081, P. R. China \\
Key Laboratory of Low Dimensional Quantum Structures and Quantum
Control (Hunan Normal University), Ministry of Education, P. R.
China.}

\author{Songbai Chen}
\email{csb3752@163.com} \affiliation{Institute of Physics and
Department of Physics,
Hunan Normal University,  Changsha, Hunan 410081, P. R. China \\
Key Laboratory of Low Dimensional Quantum Structures and Quantum
Control (Hunan Normal University), Ministry of Education, P. R.
China.}

\vspace*{0.2cm}
\begin{abstract}
\baselineskip=0.6 cm
\begin{center}
{\bf Abstract}
\end{center}

We propose two improved parameterized form for the growth index of
the linear matter perturbations: (I)
$\gamma(z)=\gamma_0+(\gamma_{\infty}-\gamma_0){z\over z+1}$ and (II)
$\gamma(z)=\gamma_0+\gamma_1
\frac{z}{z+1}+(\gamma_{\infty}-\gamma_1-\gamma_0)(
\frac{z}{z+1})^{\alpha}$. With these forms of $\gamma(z)$, we
analyze the accuracy of the approximation the growth factor $f$ by
$\Omega^{\gamma(z)}_m$ for both the $\omega$CDM model and the DGP
model. For the first improved parameterized form, we find that the
approximation accuracy is enhanced at the high redshifts for both
kinds of models, but it is not at the low redshifts. For the second
improved parameterized form, it is found that $\Omega^{\gamma(z)}_m$
approximates the growth factor $f$ very well for all redshifts. For
chosen $\alpha$, the relative error is below $0.003\%$ for the
$\Lambda$CDM model and $0.028\%$ for the DGP model when
$\Omega_{m}=0.27$. Thus, the second improved parameterized form of
$\gamma(z)$ should be useful for the high precision constraint on
the growth index of different models with the observational data.
Moreover, we also show that $\alpha$ depends on the equation of
state $\omega$ and the fractional energy density of matter
$\Omega_{m0}$, which may help us learn more information about dark
energy and DGP models.

\end{abstract}

\pacs{95.36.+x; 98.80.Es; 04.50.-h}

\maketitle

\newpage

Recently, dark energy and modified gravity have been attracted a lot
of attention because that both of them can provide a possible way to
explain the accelerating expansion of our present Universe which has
been strongly confirmed by many observations \cite{Sne, CMB, SDSS}.
In general, dark energy is regarded as an exotic energy component
with negative pressure. The modified gravity are such a kind of
theories which modify Einstein's general relativity including the
scalar-tensor theory \cite{BDT}, the $f(R)$ theory \cite{FRT} and
the Dvali-Gabadadze-Porrati (DGP) braneworld scenarios \cite{DGP}.
Since the dark energy and modified gravity can give rise to the
current accelerated expansion, it is natural to ask which one
describes correctly the real evolution of the Universe
\cite{r12,r13,r14,r15,r16,r17}. It is well known that in the same
cosmic expansion history the growth of matter perturbations are
different in the different theoretical models \cite{Starobinsky1998,
Huterer2007, Sereno2006, Knox2006, Ishak2006, Acquaviva2008,
Daniel2008, Sapone2007, Ballesteros2008, Bertschinger2008,
Laszlo2008, kunz2007, Kiakotou2008,  Linder2007, Wei2008, Wei20082,
Linder2005, Amendola,32d, Nesseris2008, Wang2008, Boisseau2000,
Gong2008, Gong2009, Polarski2008, Gannouji2008, Gannouji20082,
Fu2009, pxu1, Sb, MIS09, SLee}. Thus the growth function of the
linear matter density $\delta(z)\equiv\delta\rho_m/\rho_m$ has been
regarded as an effective tool to distinguish the dark energy and the
modified gravity at present.

At scales much smaller than the Hubble radius, the growth function
$\delta(z)$  satisfies the simple equation \cite{LWang}
\begin{eqnarray}
\label{denpert} \ddot{\delta}+2H\dot{\delta}-4\pi
G_{eff}\,\rho_m\delta=0,
\end{eqnarray}
where the dot denotes the derivative with respect to the time $t$.
$G_{eff}$ is an effective gravity constant. After defining the
growth factor $f\equiv d\ln\delta/d\ln a$, one can find that
Eq.(\ref{denpert}) becomes
\begin{eqnarray}
\label{grwthfeq1} {d\; f\over d\ln
a}+f^2+\left(\frac{\dot{H}}{H^2}+2\right)f=\frac{3}{2}\frac{G_{eff}}{G_{N}}\Omega_m.
\end{eqnarray}
where $\Omega_m=\rho_m/3H^2$ and $G_N$ is the Newton gravity
constant in general relativity. In Ref. \cite{Fry1985}, the growth
factor $f$ can be approximated very well as
\begin{equation}
\label{fommegam} f=\Omega_m^\gamma,
\end{equation}
where $\gamma$ is so-called the growth index. In general, it is a
function of redshift $z$. At the high redshift, one can set
$\Omega_m=1$ and obtain that $\gamma_\infty=\frac{3(1-w)}{5-6w}$
\cite{Linder2007, Linder2005} for the $w$CDM model and
$\gamma_\infty =11/16$ \cite{Linder2007, Wei2008} for the DGP
model~\cite{DGP}. However, at the low redshift, it is very difficult
to obtain the analytical expression of $\gamma$. Since the growth
index varies with the redshift $z$, the authors in Refs.
\cite{Polarski2008, Gannouji2008, Gannouji20082, Fu2009} proposed a
linear approximation of $\gamma(z)$, i.e., $\gamma(z)\approx
\gamma_0 + \gamma_1 z$, and found that the sign of $\gamma_1$ is
negative for $w$ CDM model and is positive for the DGP model. Thus
they claimed that the signs of $\gamma_1$ may provide another
signals to discriminate the dark energy and the modified gravity
\cite{Polarski2008, Gannouji2008, Gannouji20082, Fu2009}. However,
in Ref.\cite{pxu1}, the authors argued that the linear expansion is
only valid at the low redshift region ($z<0.5$) and then the signs
of $\gamma_1$ cannot discriminate different models from current
observations \cite{LG2008, MC2001, MT2006, NP2007, Jda2008, PM2005,
MV2004, MV2006} because that there are few growth factor data points
at $z<0.5$. Thus, the authors \cite{pxu1} proposed that the growth
index has a form
\begin{equation}
\label{gamwu} \gamma(z)=\gamma_0+\gamma_1 {z\over z+1}.
\end{equation}
\begin{figure}[htbp]
 \includegraphics[width=0.45\textwidth]{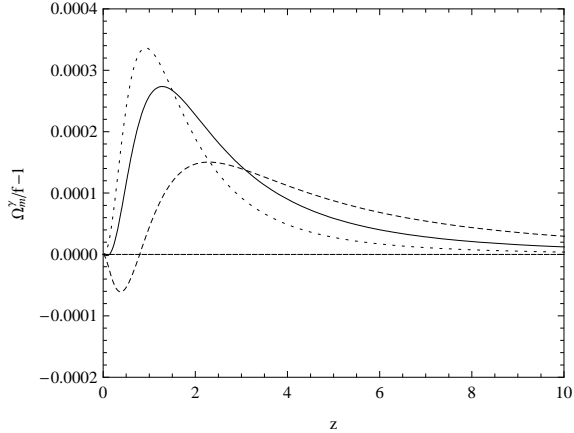}
\caption{\label{fig1} The relative difference between the growth
factor $f$ and $\Omega^{\gamma(z)}_{m}$ with redshift  for the
$w$CDM model with $\Omega_{m0}=0.27$. Here
$\gamma(z)=\gamma_0+\gamma_1 {z\over z+1}$.  The solid, dashed and
dotted curves correspond to $w=-1$, $-0.8$ and $-1.2$, respectively.
}
\end{figure}
\begin{figure}[htbp]
 \includegraphics[width=0.45\textwidth]{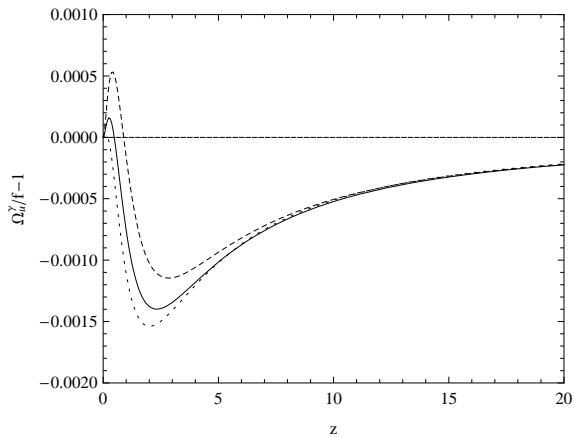}
\caption{\label{fig2} The relative difference between the growth
factor $f$ and $\Omega^{\gamma(z)}_{m}$ with the redshift  for the
DGP model. Here $\gamma(z)=\gamma_0+\gamma_1 {z\over z+1}$.  The
solid, dashed and dotted curves correspond to $\Omega_{m0}=0.27$,
$0.24$ and $0.30$, respectively.}
 \end{figure}
The merit of such a form $\gamma(z)$ is that it is applicable to all
the data points and can be used to distinguish the models using
observational data. Moreover, they also found \cite{pxu1} that this
form of $\gamma$ yields that $\Omega^{\gamma(z)}_m$ approximates the
growth factor $f$ very well both for the $\Lambda$CDM model (the
error is $\sim 0.03\%$) for all redshifts when $\Omega_{m0}=0.27$
(see fig.(\ref{fig1}) ) and for the DGP model ( the error is $\sim
0.18\%$) (see fig.(\ref{fig2})). However, it is very easy to find
from Eq.(\ref{gamwu}) that as the redshift $z\rightarrow \infty$ the
growth index $\gamma (z)$ approaches to $\gamma_0+\gamma_1$ rather
than $\gamma_{\infty}$. In the evolution of the Universe the early
difference may affect the behaviors of the growth factor $f$ at the
low redshifts. Thus, it is necessary to enhance the accuracy of the
parametrization (\ref{gamwu}) at the high redshifts.

A natural improvement to the growth index (\ref{gamwu}) is
\begin{equation}
\label{gamim1} \gamma(z)=\gamma_0+(\gamma_{\infty}-\gamma_0){z\over
z+1}.
\end{equation}
Obviously, the above growth index tends to $\gamma_{\infty}$ as
$z\rightarrow \infty$. In order to check whether our improve form of
parametrization Eq. (\ref{gamim1}) yields more accuracy than the
form (\ref{gamwu}) in Ref. \cite{pxu1},  we must evaluate the values
of $\gamma_0$ and  $\gamma_{\infty}$. The previous discussions tell
us that the expressions of $\gamma_{\infty}$ for the $w$CDM model
and the DGP model are given by \cite{Linder2007,
Linder2005,Wei2008}. For the $w$CDM model and the DGP model, the
Friedmann equations give
\begin{figure}[htbp]
 \includegraphics[width=0.45\textwidth]{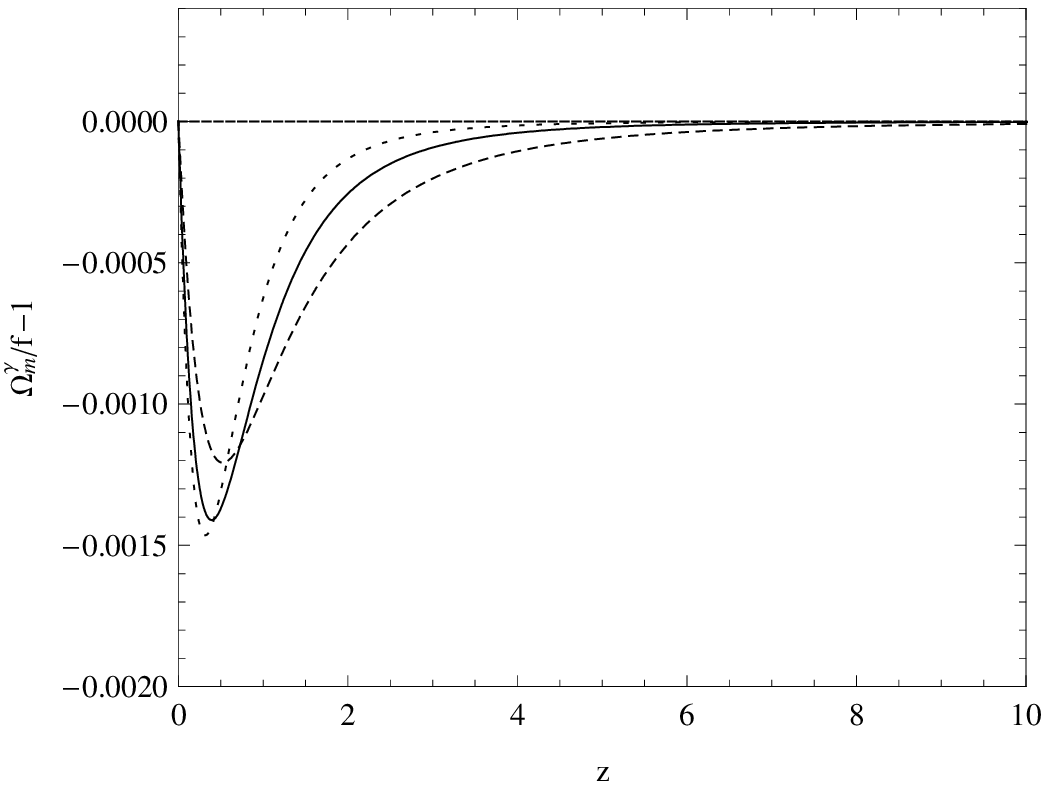}
\caption{\label{fig3} The relative error
$\frac{\Omega_m^{\gamma(z)}-f}{f}$ with redshift  for the $w$CDM
model with $\Omega_{m0}=0.27$. Here
$\gamma(z)=\gamma_0+(\gamma_{\infty}-\gamma_0) {z\over z+1}$. The
solid, dashed and dotted curves correspond to $w=-1$, $-0.8$ and
$-1.2$, respectively. }
 \end{figure}
\begin{figure}[htbp]
 \includegraphics[width=0.45\textwidth]{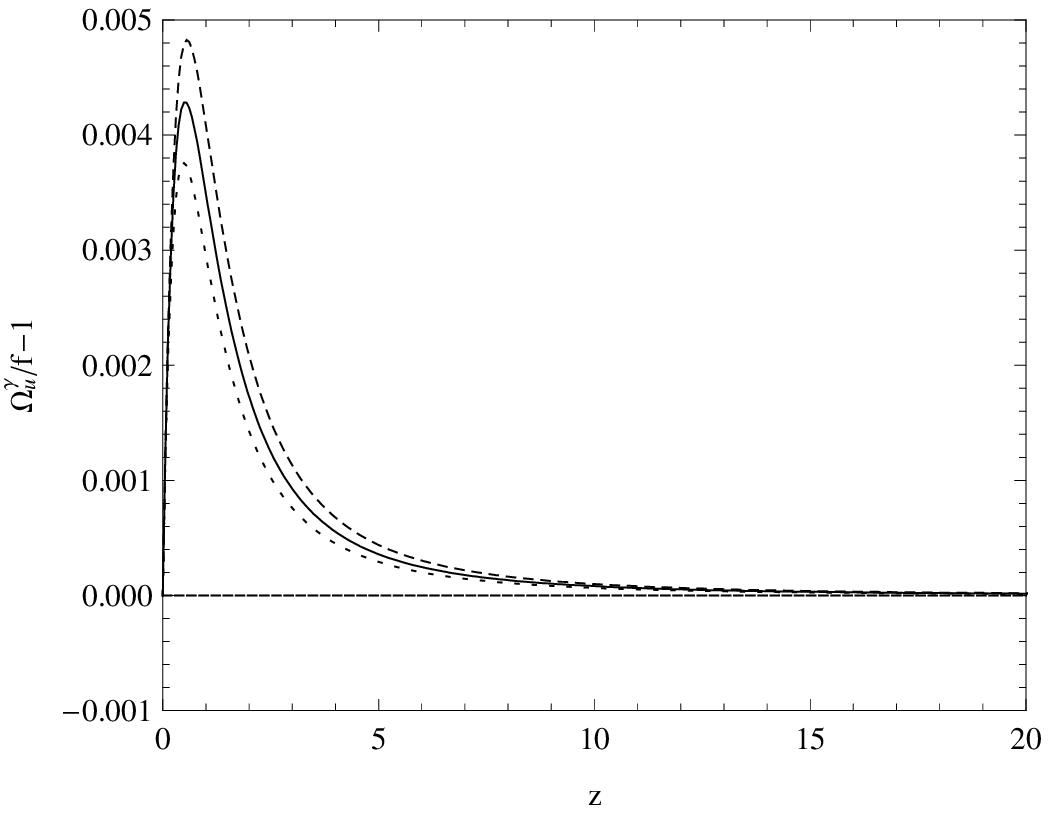}
\caption{\label{fig4} The relative error
$\frac{\Omega_m^{\gamma(z)}-f}{f}$ with redshift for the DGP model.
Here $\gamma(z)=\gamma_0+(\gamma_{\infty}-\gamma_0) {z\over z+1}$.
The solid, dashed and dotted curves correspond to
$\Omega_{m0}=0.27$, $0.24$ and  $0.30$, respectively.}
 \end{figure}
\begin{equation}
\frac{\dot{H}}{H^2}=-\frac{3}{2}w(1-\Omega_m),\;\;\;\;\;G_{eff}=G_N,\label{dkH}
\end{equation}
and
\begin{eqnarray}\label{eq1q}
 \frac{\dot{H}}{H^2}=-\frac{3\Omega_m}{1+\Omega_m},\;\;\;\;\;G_{eff}=\frac{2(1+2\Omega^2_m)}{3(1+\Omega_m^2)}
 G_{N},\label{dgpH}
 \end{eqnarray}
 respectively. Substituting Eqs. (\ref{dkH}) and (\ref{dgpH}) into Eq.
 (\ref{grwthfeq1}), we can obtain the value of $f$ at $z=0$ for
 given $\Omega_{m0}$ by resorting to numerical methods, and then get
the value of $\gamma_0$ by the relation
$\gamma_0=\ln{f(z=0)}/\ln{\Omega_{m0}}$. Through a simple
comparison, one can find that the value of $\gamma_0$ is the same as
that in Refs. \cite{Polarski2008, Gannouji2008, Gannouji20082,
Fu2009,pxu1}. Make using of the values of $\gamma_{\infty}$,
$\gamma_0$ and the improved growth index (\ref{gamim1}), we plot the
relative error $\frac{\Omega_m^{\gamma(z)}-f}{f}$ for the
$\Lambda$CDM model in fig.(\ref{fig3}) and for the DGP model in
fig.(\ref{fig4}).

Comparing figs. (\ref{fig1}), (\ref{fig2}) and (\ref{fig3}),
(\ref{fig4}),  We can obtain that for both $\omega$CDM and DGP
models the quantity $\Omega^{\gamma(z)}_{m}$ with the improved
growth index (\ref{gamim1}) approximates the growth factor $f$
better than the growth index (\ref{gamwu}) at the high redshifts.
But at the low redshifts, the relative error
$\frac{\Omega_m^{\gamma(z)}-f}{f}$ obtained by the growth index
$\gamma(z)=\gamma_0+(\gamma_{\infty}-\gamma_0){z\over z+1}$ is
larger than that by $ \gamma(z)=\gamma_0+\gamma_{1} {z\over z+1}$ in
Ref. \cite{pxu1}. Thus, the improved growth index (\ref{gamim1}) is
not good approximation for the $\gamma(z)$ in Eq. (\ref{fommegam}).

In order to make use of the virtues of the growth indexes
(\ref{gamwu}) and (\ref{gamim1}), we propose another improved
parameterized form on $\gamma(z)$
\begin{equation}
\label{gamim2} \gamma(z)=\gamma_0+\gamma_1
\frac{z}{z+1}+(\gamma_{\infty}-\gamma_1-\gamma_0)\bigg(
\frac{z}{z+1}\bigg)^{\alpha},
\end{equation}
where $\alpha$ is a numerical parameter depended on the equation of
state $\omega$ and the fractional energy density of matter
$\Omega_{m0}$.  Obviously, as $z\rightarrow 0$ and $z\rightarrow
\infty$, $\gamma(z)$ approach to $\gamma_0$ and $\gamma_{\infty}$,
respectively. Here we assume $\alpha>1$ so that the third term in
the $\gamma(z)$ can be regarded as a higher order correction to the
growth index (\ref{gamwu}). This growth index $\gamma(z)$ contains
four parameters and is more complicated than in (\ref{gamwu}).
However, it will improved the accuracy of the approximation.
Moreover, as the parameter $\gamma_1$ in (\ref{gamwu}), the
coefficient $\gamma_{\infty}-\gamma_1-\gamma_0$ and exponent
$\alpha$ in the third term can provide us more ways to understand
the differences between dark energy and DGP models.

Similarly, in order to check how well $\Omega^{\gamma(z)}_{m}$ with
the improve form of parametrization Eq. (\ref{gamim2}) approximates
the growth factor $f$, we must obtain the values of the parameters
(i.e., $\gamma_0$, $\gamma_1$ and $\gamma_{\infty}$) appeared in Eq.
(\ref{gamim2}). The calculation of $\gamma_0$ and $\gamma_{\infty}$
are similar to those in the previous discussion. As in Ref.
\cite{Polarski2008, Gannouji2008, Gannouji20082, Fu2009, pxu1}, the
value of $\gamma_1$ can be approximated by the derivative
$\gamma'(z)$ at redshift $z=0$ because that the derivative of the
third term in the $\gamma(z)$ (\ref{gamim2}) with respect to $z$ is
equal to zero at the point $z=0$ since $\alpha>1$. Thus, for the
$\omega$CDM model and the DGP model, $\gamma_1$ can be obtained by
\begin{eqnarray}
 \label{gamma-3}
 \gamma_1={1\over \ln{\Omega_{m0}^{-1}}}\bigg[{3\over 2}\Omega_{m0}^{1-\gamma_0}-\Omega_{m0}^{\gamma_0}-{3\over
 2}w
 (2\gamma_0-1)(1-\Omega_{m0})-{1\over2}\bigg]\;,
 \end{eqnarray}
 and
\begin{eqnarray}
 \label{gamma0prime-1}
  \gamma_1&=&{1\over \ln{\Omega_{m0}^{-1}}}\bigg[-\Omega_{m0}^{\gamma_0}
  +\frac{1+2\Omega_{m0}^2}{1+\Omega_{m0}^2}\Omega_{m0}^{1-\gamma_0}-{1\over
  2}+\frac{3(1-\Omega_{m0})}{1+\Omega_{m0}}
  (\gamma_0-{1 \over
  2})\bigg]\,,
 \end{eqnarray}
 respectively. Obviously, the forms of $\gamma_1$ for both models are identical
 to those in \cite{Polarski2008, Gannouji2008, Gannouji20082, Fu2009, pxu1}.
\begin{figure}[htbp]
\includegraphics[width=0.45\textwidth]{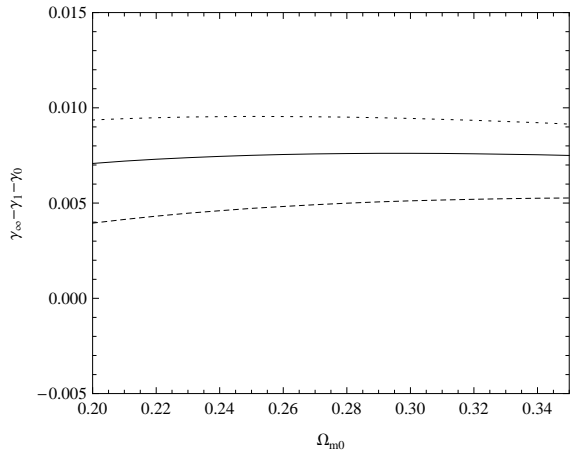}
\caption{\label{fig05} The  $\gamma_{\infty}-\gamma_1-\gamma_0$  for
$w$CDM model with $0.25\leq\Omega_{m0}\leq0.30$. The solid, dashed
and dotted curves correspond to $w=-1$, $-0.8$ and $-1.2$,
respectively. }
 \end{figure}
\begin{figure}[htbp]
\includegraphics[width=0.45\textwidth]{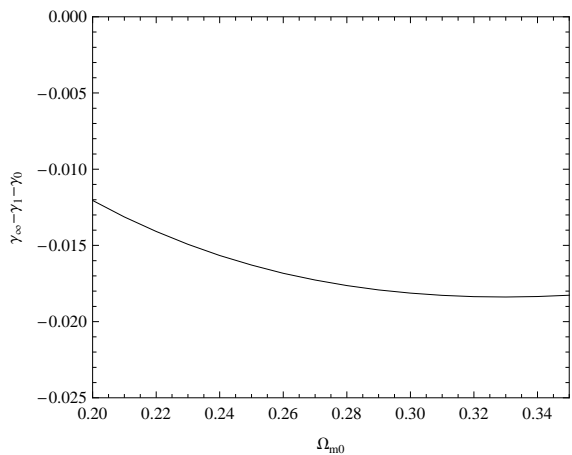}
\caption{\label{fig06} The $\gamma_{\infty}-\gamma_1-\gamma_0$ for
DGP model with $0.25\leq\Omega_{m0}\leq0.30$.}
\end{figure}
Moreover, we find that in the third term in (\ref{gamim2}) the
coefficient $\gamma_{\infty}-\gamma_1-\gamma_0$ is positive for dark
energy and is negative for DGP model, which is plotted in
Figs.(\ref{fig05})-(\ref{fig06}). The exponent $\alpha$ can be
estimated by using the value of $\gamma(z)$ at $z=z_0$
\begin{equation}
\alpha=\bigg[\ln{\frac{z_0}{z_0+1}}\bigg]^{-1}\ln{\bigg[\frac{\gamma(z_0)-\gamma_0-\gamma_1\frac{z_0}{z_0+1}}{\gamma_{\infty}-\gamma_0-\gamma_1}\bigg]}.
\end{equation}
Obviously, the above expression of $\alpha$ gives the different
values for different $z_0$. In the approximation of growth index
$\gamma(z)$ (\ref{gamwu}), the error is larger at the low redshift.
Thus, we take the value $\alpha$ at $z_0=1$ for simplicity in our
evaluation.  In Figs.(\ref{fig5})-(\ref{fig6}), we show the possible
region of $\alpha$ with a given region of $\Omega_{m0}$: $0.25
\leq\Omega_{m0}\leq0.30$. From these figures, we find that $\alpha$
depends on the equation of state $\omega$ and the fractional energy
density of matter $\Omega_{m0}$. The the coefficient
$\gamma_{\infty}-\gamma_1-\gamma_0$ and exponent $\alpha$ in the
corrected term contain more information about dark energy and DGP
models, which could provide us more methods to discriminate them.
\begin{figure}[htbp]
\includegraphics[width=0.45\textwidth]{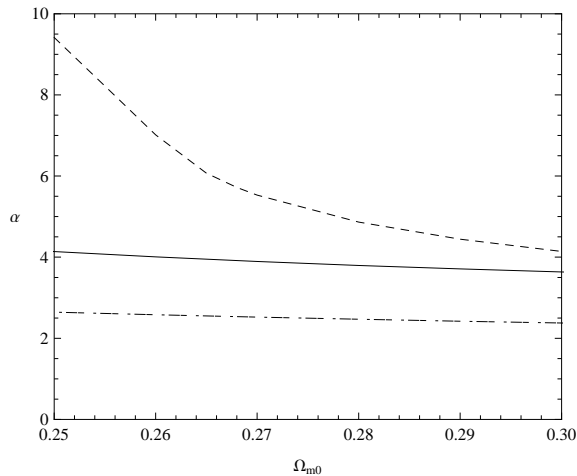}
\caption{\label{fig5} The  $\alpha$  for $w$CDM model with
$0.25\leq\Omega_{m0}\leq0.30$. The solid, dashed and dotted curves
correspond to $w=-1$, $-0.8$ and $-1.2$, respectively. }
 \end{figure}

\begin{figure}[htbp]
\includegraphics[width=0.45\textwidth]{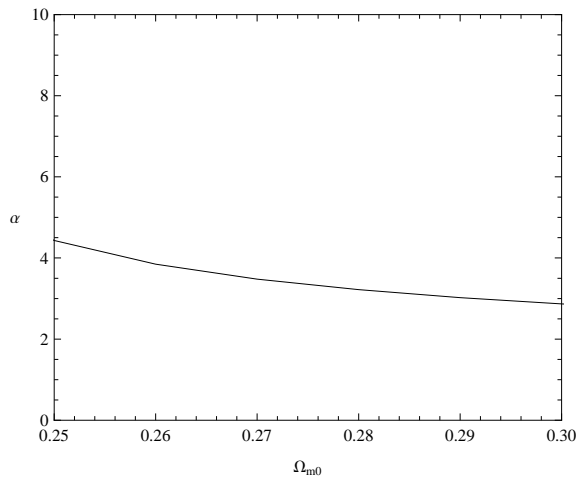}
\caption{\label{fig6} The $\alpha$ for DGP model with
$0.25\leq\Omega_{m0}\leq0.30$.}
\end{figure}

\begin{figure}[htbp]
\includegraphics[width=0.45\textwidth]{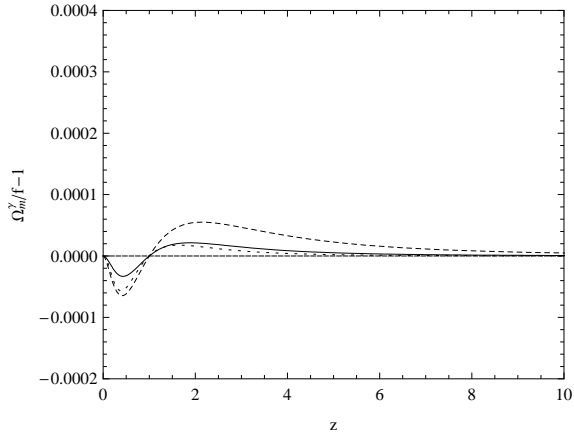}
\caption{\label{fig7} The relative error
$\frac{\Omega_m^{\gamma(z)}-f}{f}$ with redshift  for the $w$CDM
model with $\Omega_{m0}=0.27$. Here $\gamma(z)=\gamma_0+\gamma_1
\frac{z}{z+1}+(\gamma_{\infty}-\gamma_1-\gamma_0)(
\frac{z}{z+1})^{\alpha}$. The solid, dashed and dotted curves
correspond to $w=-1$, $-0.8$ and $-1.2$, respectively. }
 \end{figure}

\begin{figure}[htbp]
\includegraphics[width=0.45\textwidth]{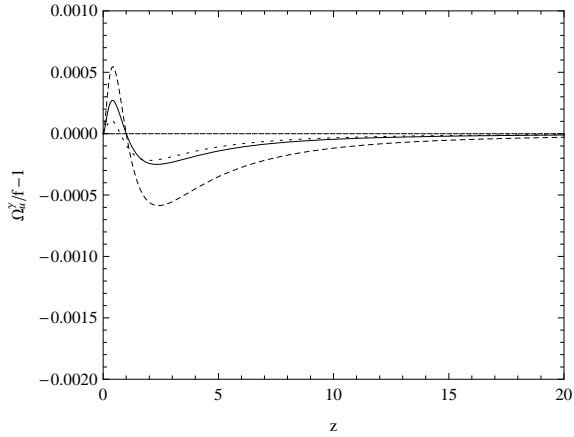}
\caption{\label{fig8} The relative error
$\frac{\Omega_m^{\gamma(z)}-f}{f}$ with redshift for the DGP model.
Here $\gamma(z)=\gamma_0+\gamma_1
\frac{z}{z+1}+(\gamma_{\infty}-\gamma_1-\gamma_0)(
\frac{z}{z+1})^{\alpha}$. The solid, dashed and dotted curves
correspond to $\Omega_{m0}=0.27, 0.24$ and $0.3$, respectively.}
 \end{figure}

Let us now adopt to the improved form of $\gamma(z)$ (\ref{gamim2})
and compare the numerical result $f$ with the analytical
approximation $\Omega^{\gamma(z)}_m$. The results for the
$\omega$CDM  and DGP models are shown in Figs. (\ref{fig7}) and
(\ref{fig8}) respectively. For the $\Lambda$CDM  model, we find that
the relative error is below $0.003\%$ for all redshifts when
$\Omega_{m0}=0.27$. This is much less than that obtained in Ref.
\cite{pxu1} where with $ \gamma(z)=\gamma_0+\gamma_{1} {z\over z+1}$
the error is only below $0.03\%$. Thus, using our improved
parameterizations of growth index (\ref{gamim2}) the error enhances
one order of magnitude improvement. Comparing Figs.(\ref{fig1}) and
(\ref{fig7}), we also find that at high redshifts
$\Omega^{\gamma(z)}_m$  approximates $f$ more accurately than that
in Ref. \cite{pxu1} for the dark energy models with different
$\omega$.

For the DGP model, one can find from fig.(\ref{fig8}) the largest
relative error $\frac{\Omega_m^{\gamma(z)}-f}{f}$ is $0.028\%$ for
all redshifts when $\Omega_{m0}=0.27$, which is also less than that
in Ref.\cite{pxu1} where the largest one is $0.18\%$.  Moreover,
comparing figs.(\ref{fig2}) and (\ref{fig8}) we find that at high
redshifts the accuracy of the approximation with our improved form
(\ref{gamim2}) is also improved  by eight times than with the old
one \cite{pxu1}. Therefore, with the second improved
parameterizations of the growth index (\ref{gamim2}), the
$\Omega^{\gamma(z)}_m$ approximates the grow factor $f$ more
accurately than those in the previous literatures
\cite{Polarski2008, Gannouji2008, Gannouji20082, Fu2009,pxu1,MIS09}
both for $\Lambda$CDM and DGP models.

In summary, we proposed two improved parameterized form for the
growth index of the linear matter perturbations and analyzed the
growth factor for both $\omega$CDM and DGP models. Using the first
improved parameterized form, we find that $\Omega^{\gamma(z)}_m$
approximates the grow factor $f$ more accurately than that in the
case the growth index $\gamma(z)$ is parameterized by
$\gamma(z)=\gamma_0+\gamma_1 {z\over z+1}$ at the high redshifts for
both kinds of models, but it is not at the low redshifts. However,
if we adopt to the second improved parameterized form, one can find
that the accuracy of the approximation the growth factor $f$ by
$\Omega^{\gamma(z)}_m$ is enhanced evidently for all redshifts. The
relative error is under $0.003\%$ for the $\Lambda$CDM model and
$0.028\%$ for the DGP model when $\Omega_{m0}=0.27$. Comparing with
those in \cite{pxu1}, such parameterizations improve almost the
approximation accuracy by one order of magnitude. Thus, the second
improved parameterized form of $\gamma(z)$ should be useful for the
high precision constraint on the growth index of different models
with the observational data.

\section*{Acknowledgments}

This work was partially supported  by the National Natural Science
Foundation of China under Grant No.10675045 and No.10875040; and the
Hunan Provincial Natural Science Foundation of China under Grant
No.08JJ3010; and the National Basic Research of China under Grant
No. 2010CB833004. S. B. Chen's work was partially supported by the
National Natural Science Foundation of China under Grant No.10875041
and the construct program of key disciplines in Hunan Province.

\end{document}